\documentclass[letterpaper]{article}
\usepackage{aaai}
\usepackage{times}
\usepackage{helvet}
\usepackage{courier}
\usepackage[linesnumbered,ruled,vlined]{algorithm2e}
\usepackage{array}
\usepackage{graphicx}
\usepackage{listings}
\begin{document}
%
\title{Solving the Subset Sum Problem\\ with Heap-Ordered Subset Trees}
\author{Daniel Shea\\
dbj6@wildcats.unh.edu\\
}
\maketitle
\begin{abstract}
\begin{quote}
In the field of algorithmic analysis, one of the more well-known exercises is the subset sum problem. That is, given a set of integers, determine whether one or more integers in the set can sum to a target value. Aside from the brute-force approach of verifying all combinations of integers, several solutions have been found, ranging from clever uses of various data structures to computationally-efficient approximation solutions. In this paper, a unique approach is discussed which builds upon the existing min-heap solution for positive integers, introducing a tree-based data structure influenced by the binomial heap. Termed the subset tree, this data structure solves the subset sum problem for all integers in time $O(N^3k\log k)$, where $N$ is the length of the set and $k$ is the index of the list of subsets that is being searched.
\end{quote}
\end{abstract}

\section{Introduction}
\noindent The subset sum problem asks whether one or more integers in a given input set sum to a target value. For example, given the set $\left\{-8, -2, 5, 7, 9\right\}$, do any of these integers sum to 10? The answer is yes, as $\left\{-2, 5, 7\right\}$ sums to 10. This may be apparent for one to see, but instructing a computer to solve a problem like this in an efficient manner may prove to be quite difficult.

The most straightforward approach to solving the subset sum problem is to test every combination of values in some systematic way. For example, we may check all integers individually to see if any are 10. If the sum is not found, then we may try all pairs of integers ($\left\{-8, -2\right\}$, $\left\{-8, 5\right\}$, ..., $\left\{7, 9\right\}$). If the sum is still not found, then we may try all triples, quadruples, and so on until we have either found the sum or exhausted all possibilities. This solution works but is inefficient for larger input sets, with a computational efficiency of $O(2^NN)$.

The approach introduced in this paper leverages a unique data structure termed a subset tree, in which all subsets of a particular length $n$ are ordered in a min-heap fashion. A binary search is then performed on this ordered set of $n$-subsets. If the target sum is not found in the subset tree of length $n$, then the subset tree of length $n+1$ is generated and searched. This process continues in an iterative fashion until the sum is found or the subsets used to generate the subset tree are equal to the length of the set. To minimize space complexity of each subset tree, the subsets are generated as needed during the binary search phase of the algorithm rather than building the complete subset tree upfront. To account for sets with negative integers when building the subset tree, each value is offset by a value that produces a set of strictly positive integers. This value is the absolute value of the minimum negative integer in the set plus one. The complexity of this algorithm is $O(N^3k\log k)$, where $N$ is the length of the set and $k$ is the index of the subset tree being located during the binary search.

A sample implementation of this algorithm has been provided\footnote{Daniel Shea. (2016). SubsetTree: Subset Tree. Zenodo. 10.5281/zenodo.51040}. The provided implementation includes sample code for each section discussed in this paper.

\section{Prior Work}
\noindent A dynamic programming solution exists in pseudo-polynomial time $O(sN)$, where $s$ is the target value and $N$ is the length of the set. This approach maintains an array of Boolean values $Q(N, s)$ and employs recursive arithmetic operations for each element in the set from the index of the target value. The value of the array at each index computed by the algorithm is set to true, indicating that there exists a subset in the set which sums to the value of the index. The solution to whether a subset exists which sums to the target value can then be found by accessing the Boolean value of $Q(N, s)$. The complexity of this algorithm has since been improved to $\tilde{O}(s \sqrt N)$ (Koiliaris and Xu, 2015).

\begin{algorithm}
\KwIn{S, t}
\KwOut{O}
Sort $S$ into increasing order\\
Define a tree $O$ where: \linebreak
 (I) The root is the singleton consisting only of the first element
\linebreak
 (II) The left child of a set whose maximum element is $i$th in the sorted input list is obtained by replacing that element by element $i+1$
\linebreak
 (III) The right child is obtained by including element $i+1$ without removing any existing element\\
Return $O$\\
\caption{Binary min-heap for subsets of positive integers}
\label{algo:a}
\end{algorithm}

Improvements to the exponential time algorithm have been explored, bringing the running time down to $O(2^{N/2})$ (Horowitz and Sahni, 1974). The algorithm takes a set of length $N$ and partitions the elements into two sorted sets of $N/2$ each. One list is stored for each of the two sets, with each list consisting of the sums for all $2^{N/2}$ possible subsets. The algorithm maintains two pointers: one pointer starting at the smallest sum in the first list and the other pointer starting at the largest sum in the second list. If the sum of these two sums equals the target value, then the search terminates. If the sum is less than the target value, then the first pointer is incremented by one position. If the sum is greater than the target value, then the second pointer is decremented by one position. This process continues until the target value is found or the search space has been exhausted.

This $O(2^{N/2})$ solution has been improved through the use of min-heaps (Schroeppel and Shamir, 1981). Rather than evaluating and sorting the list of sums for all subsets in the divided set, these sums are generated in order through the use of the min-heap data structure. The first list is divided into two additional lists $a$ and $b$, resulting in a length of  $N/4$. The subsets from each list are generated and sorted in increasing order. A min-heap is then constructed from the pairs of subsets from $a$ and $b$. As each subset from the min-heap is needed, it is popped from the top of the heap and the head element $(a_i, b_j)$ is replaced with the subset $(a_{i+1}, b_j)$. Subsequently, all pairs of subsets from $a$ and $b$ are generated in order of their total sum. Likewise, the second list is divided into two additional lists and the subsets are generated in order. However, these subsets are generated in decreasing order via a max-heap. The aforementioned exponential algorithm is then applied in which the subsets of the first list are incremented and the subsets of the second list are decremented as necessary until the target value is found or both heaps become empty. The running time of this method is $O(N2^{N/2})$.

\section{Min-Heap Solution with Positive Integers}
\noindent Building upon the prior work with min-heaps of subsets, a binary min-heap may be used to extend the solution to the initial input set as a whole when the input set contains only positive integers. A binary min-heap is a binary tree whose parent nodes are less than or equal to their child nodes. Because it makes no guarantee of total sorting (i.e. the root node is always the smallest in the tree), it is considered a partially-ordered data structure. Assuming all positive integers, Algorithm 1 is used to generate a tree of subsets in order of increasing sum, where $S$ is the input set of positive integers, $t$ is the target value, and $O$ is the subset that sums to the target value.

Figure 1 visualizes the complete min-heap binary tree for the set $\left\{2, 5, 7\right\}$. By repeating Step 2 of Algorithm 1 for each node, the heap-ordered binary tree is generated as demonstrated.

The $k$ smallest elements of the heap-ordered tree can be found in time $O(k)$ (Frederickson, 1993). This is done through the hierarchical grouping of particular elements in the heap and maintaining them in a recursively defined data structure. An alternate solution to finding the $k$th smallest element of a min-heap is shown in Algorithm 2, where the input $T$ is a min-heap binary tree of subsets. In this case, the lookup is performed in time $O(k\log k)$.

\begin{figure}
\includegraphics[width=\columnwidth]{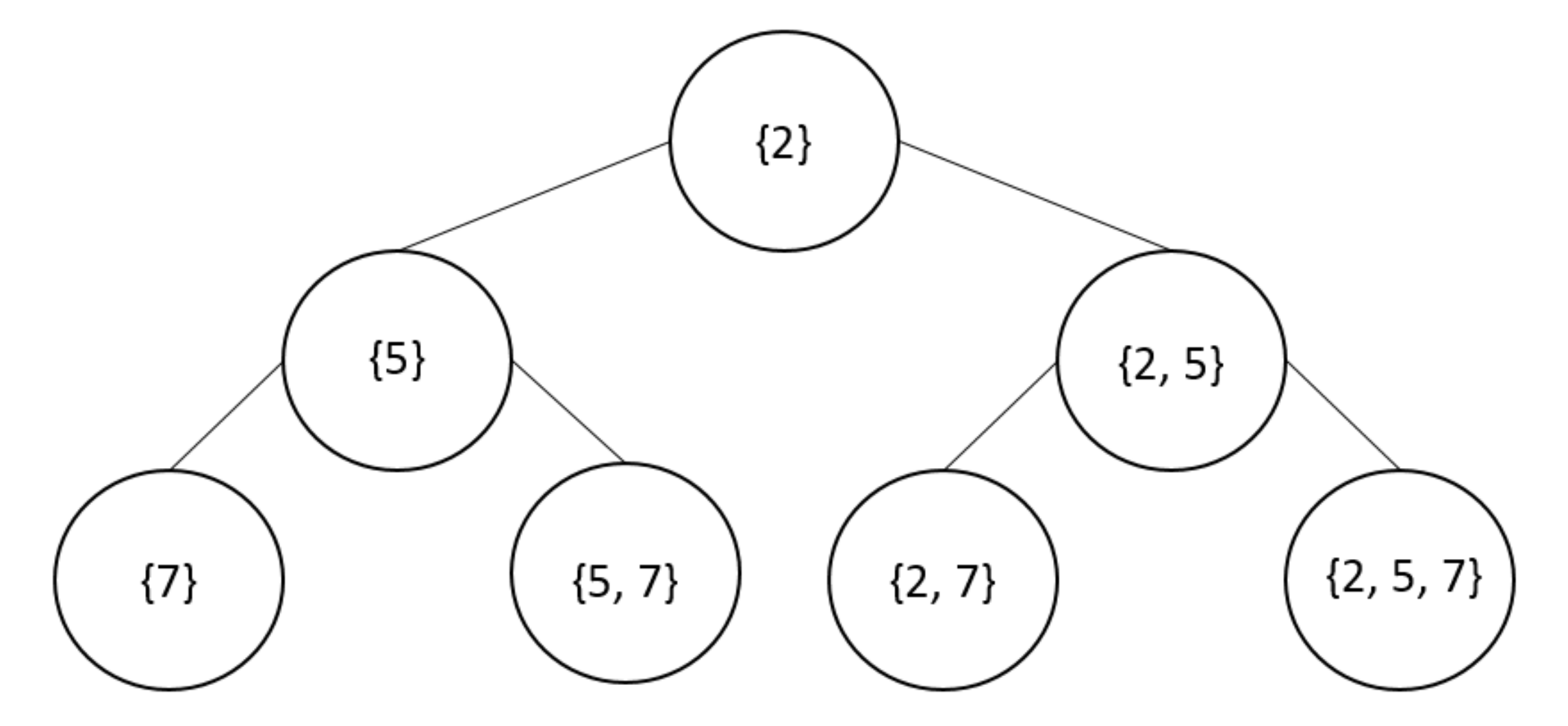}
\caption{A min-heap binary tree for $\left\{2, 5, 7\right\}$.}
\end{figure}

Because each child node is trivially derived from its parent node, the search may be performed in a lazy manner and child nodes may be generated only as required for the search. In the aforementioned example $\left\{2, 5, 7\right\}$, if the 4th smallest subset is desired, then the root node of $\left\{2\right\}$ and its children, $\left\{5\right\}$ and $\left\{2, 5\right\}$, are generated. Per Algorithm 2, the subsequent child nodes for both of the root node's child nodes are generated in the same step as adding the root node's children to \textit{M}. Thus, the lookup is effectively performed on a virtual array without the need to generate all nodes in the power set of the input.

A binary search can be performed on the sorted list of subsets that can be made from the input set. Since there are $2^N$ subsets in total, a binary search to find a target sum $t$ can be performed in time $O(\log 2^N)$. This reduces to $O(N\log 2)$, or $O(N)$. Since each search for the $k$th largest element takes $O(k\log k)$ time, a binary search on the virtual sorted list of all subsets to find the sum will take $O(Nk\log k)$ time, where $k\leq 2^N$.

However, if negative integers are included in the initial set alongside positive integers, then the approach taken to construct a heap-ordered binary tree will not suffice, as Algorithm 1 would place negative integers in the child nodes of subsets of strictly positive integers and violate the properties of the min-heap. Therefore, a modification to Algorithm 1 is necessary to handle cases of sets with both positive and negative integers.

\section{Inclusion of Negative Integers}
\noindent Assume a set of integers with both positive and negative values. To continue using a heap-ordered tree constructed in a similar fashion as previously discussed, all values in the set must be scaled by some offset value to be made positive. This can be done by adding the absolute value of the least element plus one to all elements. The offset applied to each element must also be stored for future use in the algorithm.

\begin{algorithm}
\KwIn{T}
\KwOut{M[k-1]}
Define a min-heap of nodes to visit $V$\\
Insert the root node of $T$ into $V$\\
Define an empty array $M$\\
While $Length(M) < k$: \linebreak
 (I) Pop the root node of $V$
\linebreak
 (II) Add the popped node to $M$
\linebreak
 (III) Add the popped node's children to $V$
\caption{Retrieve the $k$th smallest element from a min-heap binary tree}
\label{algo:b}
\end{algorithm}

As an example, take the set $\left\{-7, -3, -2, 5, 8\right\}$ and the target sum $0$. The scaled input is $\left\{1, 5, 6, 13, 16\right\}$. The offset value $8$ must also be stored, as it will be applied to the target sum later.

\begin{figure}
\includegraphics[width=\columnwidth]{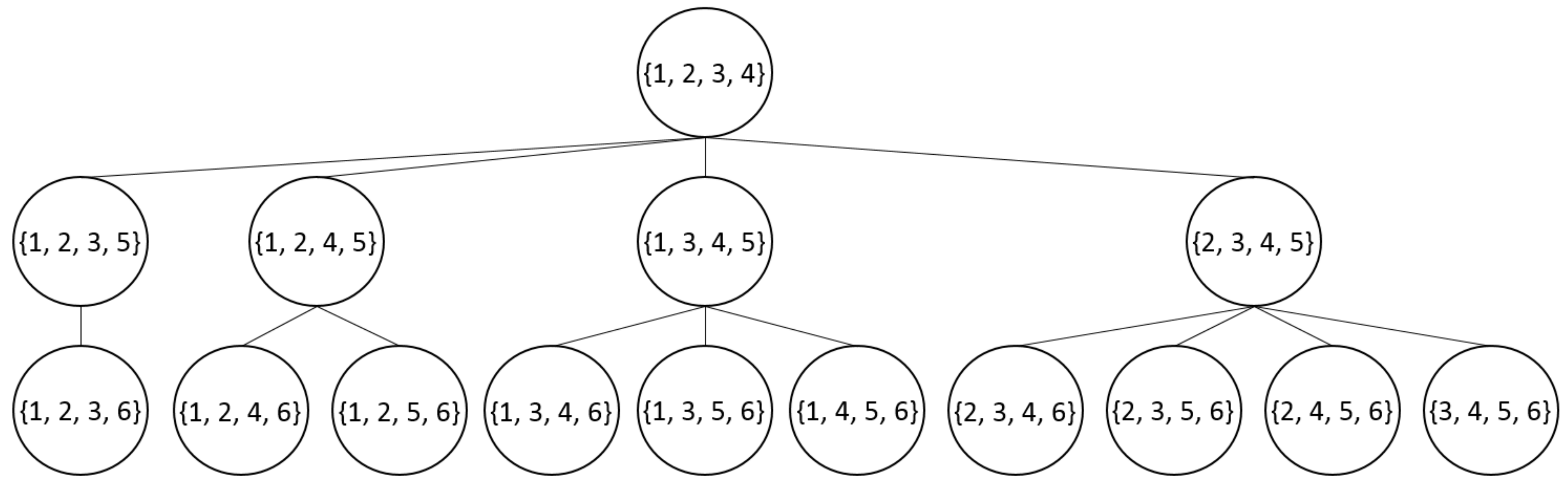}
\caption{A heap-ordered 4-subset tree for the set $\left\{1, 2, 3, 4, 5, 6\right\}$.}
\end{figure}

Despite this new set consisting of all positive values, Algorithm 1 cannot be applied as is. This is due to the lack of a consistent target value for subsets of different lengths. For example, although $\left\{-3, -2, 5\right\}$ sums to $0$, the scaled subset $\left\{5, 6, 13\right\}$ will only sum to the target value by adding the offset to the target three times. In this case, the target value must be scaled from $0$ to $24$. Generally, the target sum can only be compared to a scaled subset of size $n$ if $offset*n$ is added to the target value itself. Therefore, $N$ min-heaps must be generated, with each one consisting of all subsets of equal length. A binary search may then be performed on each min-heap.

Maintaining an ordered min-heap of subsets of the same length can be accomplished through the use of a unique data structure similar to the binomial heap, introduced here as the heap-ordered subset tree.

\section{The Heap-Ordered Subset Tree}
\noindent The subset tree is a heap-ordered tree structure consisting of $n$-length subsets of a set $S$ of length $N$. The process of constructing a subset tree $T$ of order $n$ from set $S$ is detailed in Algorithm 3.

For a subset tree of order $n$, the total number of nodes is bounded at ${N}\choose{n}$, where $N$ is the total number of elements in the input set. Each node has at most $N$ children and the height of the tree is at most $N$.

To demonstrate a simple example, Figure 2 displays the 4-subset tree for the set $\left\{1, 2, 3, 4, 5, 6\right\}$.

\begin{algorithm}
\KwIn{S, n}
\KwOut{T}
Sort the input set $S$\\
Set the root node of the subset tree $T$ to the $n$ smallest elements of $S$\\
The first child node is the subset of the parent node with the parent node's greatest element replaced with the next greatest element in $S$. If no such greater element exists, then the end of $S$ has been reached and this child node should not be generated.\\
Let the variable $i$ be equal to the penultimate index in the subset of the parent node: \linebreak
 (I) The next child node is the subset of the parent node with the element at index $i$ replaced by the next greatest element from index $i$ in $S$. If that element already exists in the subset, then increment the conflicting element with its next greatest element in $S$.
\linebreak
 (II) Repeat with any additional conflicts until none exist or there is no greater element in $S$ with which to increment. If the latter is the case, then the end of $S$ has been reached and this child node should not be generated.\\
Decrement $i$ by 1 and repeat Step 4 for all subsequent child nodes until $i$ is less than the smallest index at which the parent node incremented its subset value\\
Terminate and return $T$ when the greatest element in the subset is equal to the greatest element in $S$\\
\caption{$n$-subset tree $T$ from input set $S$}
\label{algo:c}
\end{algorithm}

As Algorithm 3 must be performed with subsets consisting of only positive elements, this operation is performed on the scaled set as opposed to the initial set.

Although this operation generates the tree in its entirety, it is trivial to generate nodes up to the $k$th smallest element due to the partial ordering in which the subset tree is constructed. This optimization minimizes the space needed for subset trees to be stored, which is especially useful for subsets of an arbitrarily large input set.

\section{Subset Sum Solution with the Subset Tree}
\noindent The complete algorithm for the subset sum problem for sets of both positive and negative integers is provided in Algorithm 4, where $S$ is a given input set of length $N$, $t$ is the target value, and $R$ is the returned subset.

The solution begins by sorting the set and uniformly scaling each member of the set to a positive value. For $1$ to $N$, where $N$ is the length of the input set, a binary search is performed on all subsets of the appropriate length. The subsets are generated in order of their sum with the use of the subset tree. This iteration is continued until a subset is found which sums to the target value or the search space has been exhausted. In the case that no subset of a given set exists for a given target value, an empty subset is returned.

\begin{algorithm}
\KwIn{S, t}
\KwOut{R}
Sort $S$\\
Offset each element in $S$ by the absolute value of the least element in $S$ plus one\\
Store this $offset$ value\\
Define a variable $O$ and set it to 1. This is the order of the virtual subset heap being searched.\\
While $O\leq N$ and a subset $R$ that sums to $S+(offset*O)$ has not been found: \linebreak
 (I) Perform a binary search for $S+(offset*O)$ on the $O$-subset tree $T$
\linebreak
 (II) If a subset $R$ is found, then terminate the loop. Else, increment $O$ by 1 and repeat Step 5(I). Do this until a subset has been found or $O > N$\\
If a subset $R$ has been found, then subtract $offset$ from each element in $R$ and return $R$. Else, return an empty subset.\\
\caption{Subset sum solution with subset trees}
\label{algo:d}
\end{algorithm}

The sorting step of the algorithm is independent to its implementation. The binary search is performed in time logarithmic to the size of the binomial coefficient ${N}\choose{n}$, where $N$ is the length of the input set and $n$ is the order of the subset tree. Each lookup is performed in time dependent on the $k$ value. Since the number of child nodes at each deleted node in the subset tree can be $N$, the heapify procedure requires additional computation, with the overhead bringing it from time $O(\log k)$ to time $O(N\log k)$; the lookup in its entirety takes $O(Nk\log k)$ time. Since this $O(N^2k\log k)$ operation must be run $N$ times in the worst case in which no subset exists, the overall complexity of this algorithm is $O(N^3k\log k)$.

\section{Algorithm Example}
\noindent Returning to the previous example of the input set $\left\{-7, -3, -2, 5, 8\right\}$ and the target value of $0$, the algorithm is executed as follows:

\begin{enumerate}
\item $\left\{-7, -3, -2, 5, 8\right\}$ is sorted. In this case, the set was already sorted.
\item An offset of $8$ is applied to $\left\{-7, -3, -2, 5, 8\right\}$, producing a scaled set of $\left\{1, 5, 6, 13, 16\right\}$.
\item $O$ is set to $1$.
\item A binary search for $S+(offset*O)$, or $8$, is performed on the subset tree of order $O$, or $1$.
\item Since no subset is found, $O$ is incremented by $1$. As $O\not> 5$, the search is continued.
\item A binary search for $S+(offset*O)$, or $16$, is performed on the subset tree of order $O$, or $2$.
\item Since no subset is found, $O$ is incremented by $1$. As $O\not> 5$, the search is continued.
\item A binary search for $S+(offset*O)$, or $24$, is performed on the subset tree of order $O$, or $3$.
\item The subset $\left\{5, 6, 13\right\}$ is found. Subtract $offset$ from each element to produce the subset $\left\{-3, -2, 5\right\}$.
\item Return $\left\{-3, -2, 5\right\}$.
\end{enumerate}

\section{Conclusion}
\noindent The subset sum problem continues to offer unique approaches to its solution through the use of various methodologies deviating from the naive brute-force approach, ranging from dynamic programming solutions to uses of data structures such as the min-heap to provide an iterative search space of subsets of positive integers. Building upon existing heap-based solutions, this paper introduced the subset tree data structure, allowing for proper heap-ordering of all subsets of an arbitrary length despite whether the input set consists of positive elements, negative elements, or both. Standard $k$th-minimum lookup procedures can then be performed through a binary search on all subsets of a particular length $n$. Iterating through all possible lengths of subsets from a given set, target values may be found in time $O(N^3k\log k)$, where $N$ is the length of the set and $k$ is the index of the list of subsets that is being searched.

Future work in this area may include improving the algorithm to allow for heap-ordering of subsets from a set of both positive and negative elements in a single tree, removing the need for the iteration step over subset trees of a particular order. This improvement would also have the auxiliary benefit of removing the need to apply and store an offset value to the initial set, further optimizing the space required for the algorithm to find a solution.

\section{References}
\begin{small}
\smallskip \noindent Frederickson, Greg N. "An optimal algorithm for selection in a min-heap." \textit{Information and Computation} 104.2 (1993): 197-214.

\smallskip \noindent Horowitz, Ellis; Sahni, Sartaj (1974), "Computing partitions with applications to the knapsack problem", \textit{Journal of the Association for Computing Machinery} 21: 277–292.

\smallskip \noindent Koiliaris, Konstantinos; Xu, Chao (2015-07-08). "A Faster Pseudopolynomial Time Algorithm for Subset Sum". arXiv:1507.02318.

\smallskip \noindent Schroeppel, Richard, and Adi Shamir. "A $T=O(2^n/2), S=O(2^n/4)$ Algorithm for Certain NP-Complete Problems." \textit{SIAM journal on Computing} 10.3 (1981): 456-464.
\end{small}
\end{document}